# Geometric Optimization of Patterned Conductive Polymer Composite-based Strain Sensors Toward Enhanced Sensing Performance

Jia-Chen Shang[a, b] [*]

[a] *School of Transportation Science and Engineering, Beihang University, 37 Xueyuan Road, Haidian District, Beijing, 100191, PR China*

[b] *Beijing Key Laboratory for High-efficient Power Transmission and System Control of New Energy Resource Vehicle, Beijing 100191, China*

[*]*Corresponding author, E-mail: by2213106@buaa.edu.cn*

**Highlights:**

- Developing and validating a novel universal piezoresistive model for flexible strain sensor.

- Capturing the conductivity tensor and piezoresistive behavior of anisotropic composites under large deformations.

- Presenting a numerical method for predicting electromechanical behavior of flexible strain sensor.

- Demonstrating a rapid and cost-effective workflow for optimizing patterned flexible strain sensor.

# 0. Abstract

The patterned design of flexible sensors enables customized performance to meet diverse application demands. However, when multiple geometric parameters and sensing metrics are involved, experimental approaches to establish structure-performance relationships become costly and inefficient. Here, a novel universal piezoresistive model—overcoming limitations of commonly used models that are only applicable to small strains and linear responses—is developed to capture the relationship between conductivity tensor components and strain. A numerical method incorporating this model simulates the electromechanical properties of conductive composites and predicts patterned strain sensors behavior. To validate this approach, a flexible strain sensor based on laser-induced graphene technology is fabricated and tested. Additionally, a rapid, cost-effective workflow combining Latin hypercube sampling and Pareto-optimal solutions is demonstrated for multi-parameter and multi-objective optimization of the sinusoidal-patterned sensor. This study provides valuable insights for investigating the structure-performance relationship of strain sensors and advances optimization methods for sensor designs.

# 1. Introduction

Flexible strain sensors have attracted significant attention in recent years due to their remarkable sensitivity [1-3], wide monitoring range [4-6], conformability [7-9], and lightweight nature [10-12], making them ideal for applications such as motion and health monitoring [13-15], structural health monitoring [16, 17], and self-sensing devices [18, 19]. To further leverage the advantages of flexible strain sensors in a broader range of applications, it is essential to enhance their sensing performance [4] while ensuring the ability to customize their performance to meet diverse sensing requirements across various applications [20, 21].

Methods for improving the sensing performance of flexible strain sensors can generally be divided into two categories: material modification [22-24] and structural innovation [25-27]. Among these, structural design approaches are simpler, more direct, and environmentally friendly [25]. For example, the pre-strain method [26], which induces microstructures in sensors, is straightforward and easy to implement. This technique can generate wrinkles [28-30], fish-scale-like structures [31] or engineering defects such as microcracks and fractures [32, 33] in the conductive layer, significantly enhancing sensor performance. However, as a macroscopic approach, pre-straining can only guide the general formation of structures and lacks precise control. The randomness of the induced structures leads to high variability in sensor performance. Sandpaper is inexpensive, readily available, and features a random distribution of gravel height and diameter [34]. Consequently, many researchers use sandpaper as a template to fabricate flexible strain sensors [19, 35, 36] that primarily rely on changes in contact resistance for their electromechanical response. These sensors exhibit high sensitivity and a wide working range [37]. Nevertheless, the limited variety of sandpaper types constrains the customizability of sensor performance. In recent years, 3D printing [20, 38], screen printing [39, 40], inkjet printing [41, 42], and laser direct writing [43-45] have been increasingly employed in the fabrication of flexible sensors. These methods are primarily influenced by tool parameters—such as equipment power and fabrication speed—while environmental and human factors have minimal impact, ensuring consistent sensor performance. Furthermore, with computer assistance, researchers can flexibly design sensor structures to optimize sensing performance for specific applications. Consequently, exploring the structure-performance relationships of sensors and designing optimal structures or patterns has become a key research focus in the field.

High stability is essential for sensors, and the key to achieving it lies in minimizing strain under load. Inspired by traditional Japanese paper-cutting art, Xu et al. [46] employed laser cutting to create kirigami structures, significantly reducing strain during stretching and enabling the sensor platform to maintain stable performance even after 60,000 tensile cycles, with almost no degradation. To address diverse strain measurement needs, Yang et al. [47] used screen printing to fabricate patterned flexible strain sensors capable of measuring unidirectional, bidirectional, omnidirectional strains, as well as principal strain and its direction. However, the relationship between the geometric parameters of these structures and their performance was not explored in the aforementioned studies. To adjust strain in the sensing layer, Sheng's team [5] used laser engraving to create grooves and punched circular holes in a composite conductive film, investigating how the size and arrangement of these features affected sensor performance. Peng et al. [48] used laser cutting to create porous microstructure, such as circular holes and elongated slots, and established relationships between geometric parameters, such as hole diameter or slot angle, and sensor performance. G. Arana et al. [49] developed flexible strain gauges by depositing carbon nanotubes on a polyimide substrate and investigated the effects of geometric parameters, such as the end loops and grid width, on sensing performance, finding a 13% increase in gauge factor (GF) after parameters adjustment. However, these studies are limited by the small number of samples, often yielding only qualitative relationships or imprecise quantitative relationships between geometric parameters and performance, thus limiting their reference value. This limitation arises from the time-consuming nature of sensor fabrication and performance testing. When numerous geometric parameters are involved, the scale of possible combinations becomes exceedingly large. Accurately establishing structure-performance relationships and identifying optimal geometric parameters for specific applications requires significant time and cost.

With the rapid advancement of Finite Element Method (FEM) technology, researchers can now simulate complex physical phenomena through numerical calculations, drawing conclusions that previously required experimental validation, thus significantly reducing research time and costs. Numerous studies have applied FEM to predict the electromechanical behavior of flexible conductive composites [50-53]. However, these studies typically focus on the piezoresistive behavior of materials without considering macroscopic structural effects, providing limited guidance for the geometric design of strain sensors. Some researchers [54] have developed multiscale models that account for geometric effects to analyze how geometric parameters influence sensor performance. Nevertheless, the representative volume elements (RVEs) used in these models are often idealized as homogeneous, while the conductive network structure of conductive composites is highly complex [55], limiting the application of these studies.

In this study, we introduced an innovative and efficient approach for optimizing the geometric parameters of flexible strain sensors using FEA. This method addresses the high costs and low efficiencies associated with traditional experimental techniques for multi-parameter design, enabling sensor designers to rapidly and cost-effectively optimize sensor patterns and achieve superior overall sensing performance. A flexible strain sensor was fabricated by infiltrating polydimethylsiloxane (PDMS) into laser induced graphene (LIG) generated from laser irradiated PI (polyimide) tape (PDMS@LIG). After testing and characterizing its electromechanical behavior, we proposed a novel, universal piezoresistive model to describe the sensor's response. The steady-state thermomechanical coupling module in ABAQUS was employed to simulate the sensor's electromechanical response. The agreement between FEA predictions and experimental results confirmed the accuracy of the piezoresistive model. Finally, numerous geometric parameters were generated, and the electromechanical performance of the corresponding sensors was simulated and analyzed to determine the optimal geometric parameters.

## 2. Experimental

### 2.1 Materials

The precursor used to generate LIG was 0.055 mm thick PI tape. PDMS (Sylgard 184) was purchased from Dow Corning, and 0.02 mm diameter copper wire was used as the connecting wire. Conductive silicone adhesive (TH-7003, SHENZHEN WATIHE TECHNOLOGY CO.,LTD, China) was applied to bond the wire to the PDMS@LIG. A Silicone-based conductive adhesive was selected because its mechanical properties after curing closely match those of PDMS, minimizing any interference with the experimental results.

**2.2 Fabrication of LIG**

Laser processing for fabricating LIG was conducted under ambient conditions. First, the adhesive side of the PI tape was adhered to the polytetrafluoroethylene (PTFE) mold. A 5 W semiconductor laser (wavelength: 450 nm, spot size: 0.05 mm) was then used to irradiate the PI tape, with a fixed laser scanning path spacing of 0.05 mm. By adjusting the laser scanning speed and power, LIG with varying electromechanical behaviors was produced. The effects of these parameters on the morphology and electromechanical properties of the resulting LIG were evaluated, and the optimal laser parameters were selected for sample fabrication in this study.

**2.3 Sensor Fabrication**

Once the LIG was generated, a liquid PDMS precursor mixture with a 10:1 base-to-curing-agent weight ratio was poured into the PTFE mold and placed under vacuum for 30 minutes. This step was essential for removing air bubbles from the PDMS and ensuring complete infiltration of the porous LIG structure by the PDMS. The mold was then placed in an air-blowing oven at 100 °C for 3 hours to cure the PDMS. After curing, the PDMS@LIG composites was removed from the mold, and the excess PI tape was carefully peeled away. Finally, conductive silicone adhesive was applied to the electrode positions on the LIG pattern to bond 0.02 mm diameter copper wires, which were used to collect electrical signals during the subsequent electromechanical tests.

The strain distribution remains uniform in the narrow portion of the dumbbell-shaped specimen during stretching, allowing for more accurate measurements of the material's mechanical and electromechanical behavior. A type 2 dumbbell-shaped standard tensile specimen with a thickness of 1.2 mm was prepared according to ISO 37:2017(E), with the LIG fully positioned within the narrow portion of the specimen. Additionally, a rectangular mold with dimensions of 40 × 22 × 1 mm was used to prepare the patterned PDMS@LIG specimens. A schematic of the dumbbell-shaped specimen fabrication process is illustrated in Fig. 1, while that of patterned specimen is shown in Fig. S1.

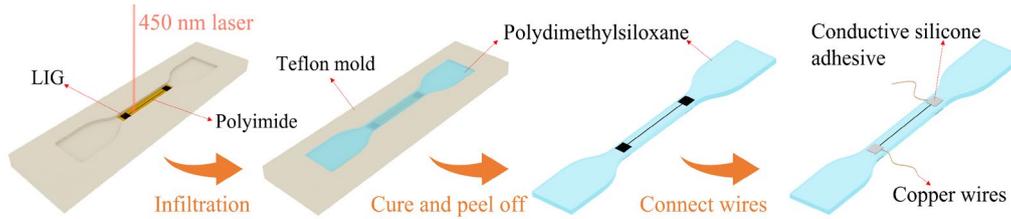

Fig. 1. Fabrication diagrams of the dumbbell-shaped PDMS@LIG specimen

## 2.4 Characterization

Scanning electron microscopy (SEM) (Apero S LoVac, Thermo Fisher Scientific) was used to characterize the microstructure of the LIG on PI tape and the PDMS@LIG composites. Confocal Raman spectroscopy (LabRAM HR Evolution, HORIBA FRANCE) with a 532 nm laser wavelength was employed for further characterization.

## 2.5 Mechanical and electromechanical performance measurement

Mechanical and electromechanical performance tests were conducted using a universal testing machine (Instron 5966) in a controlled laboratory environment at a tensile speed of 10 mm/min. Electrical resistance changes during the tests were recorded with a Keysight 34465A digital multimeter (Santa Rosa CA, USA). A DC regulated power supply (UTP3305, Uni-trend, China) provided constant current during resistivity measurements. To accurately measure the strain in the narrow portion of the dumbbell-shaped specimen, an Instron-8801 test system equipped with a video extensometer was used to retest the specimen under the same tensile speed and displacement in the same laboratory environment. It was assumed that the specimen's mechanical behavior remained identical in both experiments.

## 3. Results and discussions

### 3.1 Characterization

After the PI tape was irradiated with the laser to form the LIG, scanning electron microscopy (SEM) was immediately used to characterize its microstructure, as shown in Figs. 2(a–c). Clear parallel laser-engraved traces can be observed in Fig. 2(a). This occurs because the laser spot is only 0.05 mm. When the dimensions of the generated pattern exceed this size, the pattern is divided into parallel laser scanning tracks. This feature leads to the anisotropic conductivity and electromechanical behavior of the LIG. A scaled schematic of the laser spot and scanning trace is provided in the Fig. 2(a), showing that the width and spacing of the laser traces correspond to the laser's parameters. Fig. 2(b) and the inset at high magnification clearly show the three-dimensional porous structure of the LIG. The cross-sectional view in Fig. 2(c) reveals a loose, porous, and fuzzy structure with an average height of approximately 30 μm. The inset presents a high-magnification image, further confirming these structural features. These porous structures allow elastic materials like PDMS to easily penetrate, facilitating the transfer of the LIG. SEM images of the PDMS@LIG conductive composites, formed by infiltrating the LIG with PDMS and curing it, are shown in Figs. 2(d–e). Fig. 2(d) provides a top view, revealing surface roughness resulting from the laser engraving tracks, microspherical structures formed by exuded PDMS, and cavities where PDMS did not fully infiltrate. These features are more clearly visible in the magnified image in Fig. 2(e). In this image, green arrows indicate PDMS microspheres and regions with full PDMS penetration, while red arrows mark cavities where PDMS did not fully penetrate. Fig. 2(f) presents an enlarged view of a cavity, revealing the porous structure of the LIG (red arrows) and PDMS (green arrows) partially covering the LIG, forming a thin film. Although this increases surface roughness, it enhances adhesion of the conductive adhesive. Additionally, the cavities where PDMS has not fully penetrated serve as contact points, enabling direct electrical connections between the conductive adhesive and PDMS@LIG composites for electrical measurements.

Besides using SEM for qualitative analysis of the microstructure, Raman spectroscopy was employed to further investigate the chemical structure. Fig. 2(g) presents the Raman spectra of LIG prepared with three different laser fluences. Laser fluence was used to label the samples instead of specific laser parameters like power and scanning speed because, firstly, many studies have indicated that laser fluence is the key factor affecting the morphology and performance of LIG [56-58], and secondly, laser parameters can differ across devices, making laser fluence a more consistent labeling method. Laser fluence $F$ can be calculated using the following formula:[1]

$$F = \frac{P}{v \cdot d} \tag{1}$$

Where, $P$ represents the average laser power, $v$ the laser scanning speed, and $d$ the laser spot diameter. The laser fluence used in this study, along with the corresponding laser parameters, is listed in Table S1. The LIG prepared with all three laser fluences exhibited characteristic peaks at 1345, 1586, 2702 cm$^{-1}$, corresponding to the D-band, G-band, and 2D-band of graphene, respectively. The relative intensity ratios of these bands provide key information about the characteristics of the LIG [59]. Specifically, the intensity ratio of the G peak to the D peak ($I_G/I_D$) reflects the defect level in graphene, with a higher $I_G/I_D$ indicating fewer defects [60]. The laser fluences of 12.62, 15.82, and 17.0 J/cm² correspond to $I_G/I_D$ ratios of 0.96, 1.38, and 1.41, respectively, suggesting that the defect level in LIG decreases as laser fluence increases. The relative intensity ratio of $I_{2D}/I_G$ provides an indication of the number of graphene layers [61]. For 12.62 J/cm², the $I_{2D}/I_G$ ratio is 1.165, indicating approximately two layers of graphene. For 15.82 and 17.0 J/cm², the $I_{2D}/I_G$ ratios are 0.812 and 0.811, suggesting the presence of approximately three layers. Fig. 2(h) shows the Raman spectrum of PDMS@LIG composites prepared with LIG at a laser fluence of 15.82 J/cm². Compared to pure LIG, the $I_G/I_D$ ratio decreased from 1.38 to 0.34, indicating that the penetration of PDMS increased disorder within the LIG structure. The intensity of the 2D peak also weakened significantly, with the $I_{2D}/I_G$ ratio dropping from 0.812 to 0.106. This change can be attributed to the fact that the LIG, originally located at the bottom of the PI tape, becomes the outermost layer of the PDMS@LIG composite after transfer [62]. Fig. 2(i) presents a line graph depicting the variation of the peak intensity ratio with laser fluence, while the inset shows a bar chart comparing the peak intensity ratios for LIG and PDMS@LIG composites.

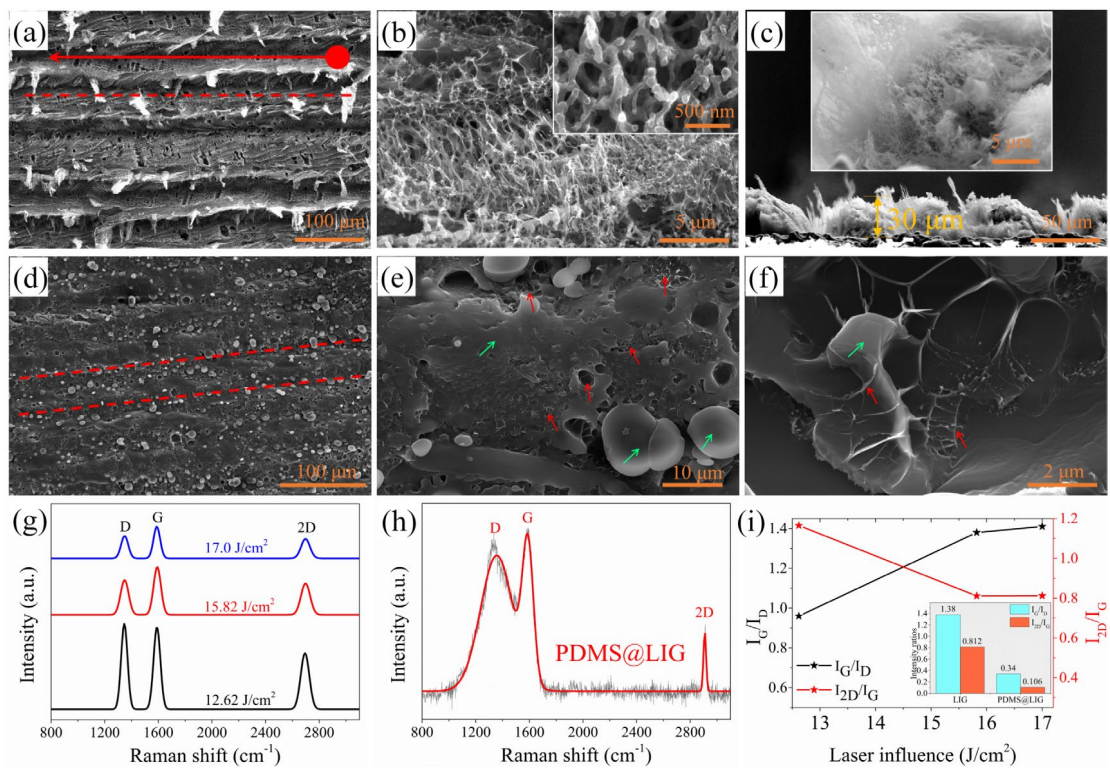

Fig. 2. Surface microstructure and Raman spectra. (a–c) SEM images of LIG; (d–f) SEM images of PDMS@LIG composites; (g) Raman spectra of LIG at different laser fluences; (h) Raman spectrum of PDMS@LIG composites; (i) Raman characteristic peak intensity ratios.

**3.2 Mechanical performance**

Mechanical properties are fundamental metrics for evaluating sensor performance and directly influence the electromechanical behavior of sensors. Accurately determining these properties is crucial for ensuring the precision of subsequent finite element simulations. To obtain these properties, type 2 dumbbell-shaped specimens prepared according to ISO 37:2017(E) was subjected to a tensile displacement of 15 mm at a speed of 10 mm/min. A video extensometer captured the axial strain changes in the narrow portion during the process, as shown in Fig. 3(a). Throughout the test, axial strain maintained a linear relationship with time, reaching 35.8% at maximum displacement.

The PDMS@LIG sensor consists of three materials: pure PDMS, PDMS@LIG composites, and conductive silicone adhesive. Due to challenges in sample preparation, accurately determining the mechanical properties of the PDMS@LIG composites and conductive adhesive is difficult. However, the mechanical behavior of a pure PDMS dumbbell-shaped specimen and a PDMS@LIG specimen with wires bonded using conductive adhesive were tested and compared. Five samples of each type were tested, and the average stress-strain values with corresponding errors are shown in Fig. 3(b), indicate a high degree of overlap. The weighted mean squared error (WMSE) between the two curves was 2.71, suggesting minimal differences. Therefore, for simplicity in finite element modeling, the mechanical properties of the three materials were treated as identical. The mean stress–strain data were fitted using a first-order incompressible Ogden model for uniaxial tension, yielding coefficients $\mu_1 = 619.660$ kPa and $\alpha_1 = 3.723$.

The PDMS@LIG specimen underwent 100 cycles of tensile with a 10 mm cross-head displacement (corresponding to an axial strain of 23.8%) at a speed of 100 mm/min. The mechanical hysteresis remained stable, with the stress at the maximum strain decreasing by approximately 4.07%. Given the large applied strain, the mechanical cycling performance can be considered reliable. The stress–strain curves of cyclic tensile tests are shown in Fig. 3(c).

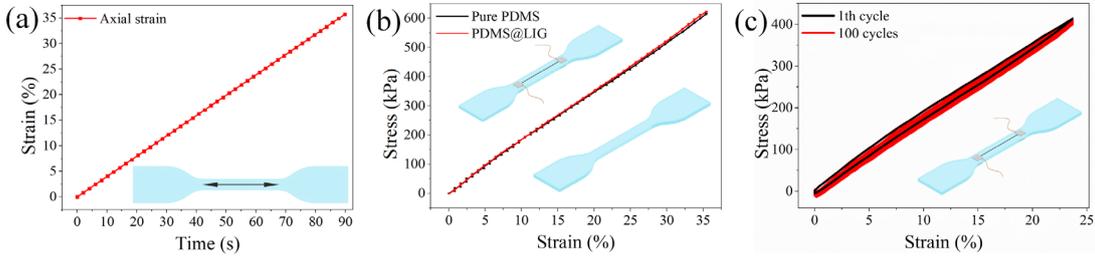

Fig. 3. Mechanical properties of the sensors. (a) Axial strain in the narrow portion recorded with a video extensometer. (b) Stress–strain curves of the pure PDMS specimen and the PDMS@LIG specimen with wires bonded using conductive adhesive. (c) Stress–strain curves of the PDMS@LIG specimen over 100 tensile loading cycles.

### 3.3 Electromechanical performance

The impact of the laser fluence on the electromechanical behavior of PDMS@LIG composites was investigated. We found that the minimum laser fluence required to generate effective LIG is 5.7 J/cm², which aligns with the threshold of 5.5 J/cm² reported in reference [63]. However, when the laser fluence exceeds 19.56 J/cm², the excessive laser power causes the PI tape to burn. Therefore, eight evenly spaced laser fluences were selected within the range of 5.7 to 19.56 J/cm². Five dumbbell-shaped specimens were prepared for each laser fluence. The average strain-normalized resistance changes and corresponding errors for these PDMS@LIG specimens are shown in Figs 4(a, b). It is evident that as the laser power increases, the sensitivity of the sensor gradually improves. However, when the laser power exceeds 17.0 J/cm², the electromechanical behavior becomes highly unstable due to damage to the LIG structure from excessive laser power. Comparing the electromechanical performance at 17.0 J/cm² and 15.82 J/cm², the normalized resistance response at maximum strain is 6.74 for 17.0 J/cm², slightly higher than 5.38 for 15.82 J/cm². However, the maximum error at 17.0 J/cm² (14.4%) is more than double that at 15.82 J/cm² (6.9%). Raman spectroscopy indicates that the degree of LIG defects at 17.0 J/cm² is similar to that at 15.82 J/cm². Considering all factors, 15.82 J/cm² was selected for LIG preparation. The inset of Fig. 4(a) presents a schematic of the specimen with a linear LIG pattern, where the LIG is aligned parallel to the tensile strain direction, referred to as parallel specimens. In contrast, specimens with LIG perpendicular to the axial strain are called perpendicular specimens. This study did not investigate the effect of the angle between the laser trace direction and the axial strain direction, as all specimens had laser traces parallel to the tensile direction, except those used for piezoresistive response measurements in the thickness direction. Additionally, the effect of LIG line width was not examined, since all LIG patterns had a consistent line width of 0.1 mm. Although some researchers have indicated that the dimensions of linear conductive composites do not significantly affect the normalized resistance response [40].

    The electromechanical responses of the perpendicular specimen were measured, with the results for both perpendicular and parallel specimens presented in Fig. 4(c). The electromechanical responses of two types of perpendicular specimens are shown, with insets indicating the corresponding specimen type. For specimen II, the laser scribing direction is parallel to the axial strain. For specimen III, the laser scribing direction is perpendicular to the axial strain and is used to measure the piezoresistive

response in the thickness direction, which will be discussed in detail later. To further clarify the preparation methods for these three specimen types, the relationship between the scribing direction and the specimens is illustrated in Fig. S2.

Electromechanical responses can be quantified using the gauge factor (GF), defined as:

$$GF = \frac{\Delta R / R_0}{\varepsilon} \tag{2}$$

Where, $\Delta R = R - R_0$ is the change in resistance, $R_0$ is the initial resistance, and $\varepsilon$ is the applied axial strain. Under an applied axial strain of 23.8%, the calculated GF is 22.6 for the parallel specimen and 1.58 for the perpendicular specimen II. After establishing the relationship between resistance changes and tensile strain in both directions, a piezoresistivity analysis is necessary to determine how conductivity varies with strain. This analysis will serve as the foundation for subsequent finite element simulations of electromechanical behavior.

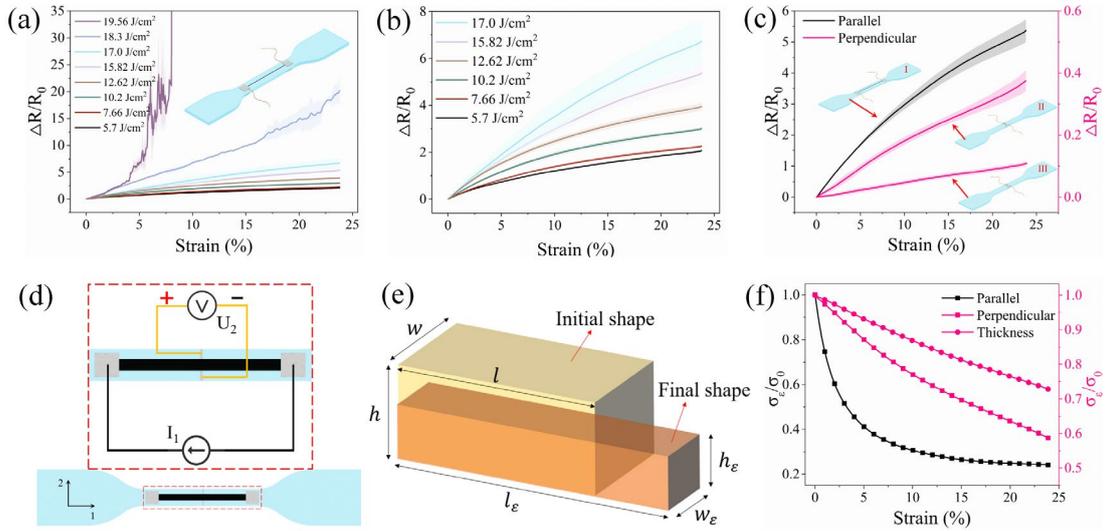

Fig. 4. Electromechanical performances of the sensors. (a, b) Influence of laser fluence on the sensor's electromechanical response. (c) Electromechanical responses of parallel and perpendicular specimens. (d) Measurement of the off-diagonal components of the resistivity tensor (e) Schematic representation of the structural changes in an elastic resistive block under strain. (f) Variation of the diagonal components of the resistivity tensor with strain.

## 3.5 Piezoresistivity analysis

Due to the anisotropy of the LIG structure, PDMS@LIG is an anisotropic conductive composite. Thus, Ohm's law in differential form is expressed as:

$$\mathbf{E} = \boldsymbol{\rho} \cdot \mathbf{j} \quad \text{or} \quad \mathbf{j} = \boldsymbol{\sigma} \cdot \mathbf{E} \tag{3}$$

Where, $\mathbf{E}$ and $\mathbf{j}$ represents the electric field intensity vector and current density vector, respectively, while $\boldsymbol{\rho}$ and $\boldsymbol{\sigma}$ are the second-order resistivity and conductivity tensors, with components $\rho_{ij}$ and $\sigma_{ij}$ ($i, j = 1, 2, 3$). According to the Casimir reciprocity relation in irreversible thermodynamics and Onsager's principle, $\boldsymbol{\rho}$ and $\boldsymbol{\sigma}$ are second-order symmetric tensors, where $\boldsymbol{\sigma} \cdot \boldsymbol{\rho} = \boldsymbol{\rho} \cdot \boldsymbol{\sigma} = \mathbf{I}$. In an orthogonal curvilinear coordinate system $\{o - u_1 u_2 u_3\}$, assuming that all quantities are functions of the coordinates, Eq. (3) can be expressed in matrix form:

$$[E_n] = [\rho_{nm}][j_m] \tag{4a}$$

$$[j_n] = [\sigma_{nm}][E_m] \tag{4b}$$

For the resistivity tensor $\boldsymbol{\rho}$, the diagonal components represent the resistivity in each principal direction, while the off-diagonal components describe the influence of current in one direction on the electric field intensity in another. For instance, $\rho_{21}$ describes the relationship between the current in the 1-direction and electric field intensity in the 2-direction. During the electromechanical response tests of the parallel and perpendicular specimens (Fig. 4(c)), the initial resistivity in the 1- directions ($\rho_{11}$) and 2-directions ($\rho_{22}$) was also obtained. The off-diagonal components of the resistivity tensor can be determined through the experiment illustrated in Fig. 4(d), where a constant current $I_1$ is applied in the 1-direction, and the potential differences $U_2$ in the 2-directions are measured. Since the current density in the 2- and 3-direction is zero, according to Eq. (4a), the electric field intensity in the 2-direction can be

expressed as:

$$E_2 = \rho_{21} j_1 \tag{5}$$

Where, $E_2 = U_2/w$, and $w$ is width of the PDMS@LIG composites, $j_1$ represents current density in the 1-direction, $j_1 = I_1/A$, $A$ is the cross-sectional area of the PDMS@LIG composites. Therefore, the resistivity $\rho_{21}$, representing the coupling effect between the 1- and 2-directions, can be calculated as:

$$\rho_{21} = \frac{E_2}{j_1} \tag{6}$$

Measurements yielded the following results: $\rho_{11} = 12.367\ \Omega\cdot\text{mm}$, $\rho_{22} = 17.751\ \Omega\cdot\text{mm}$ and $\rho_{12} = \rho_{21} = -1.14\ \Omega\cdot\text{mm}$. It is important to note that, due to the thinness of the PDMS@LIG composites, the electric field intensity in the thickness direction (3-direction) cannot be experimentally measured. However, based on the laser fabrication method, both the 1- and 3-directions exhibit a continuous porous structure. Therefore, it can be assumed that the PDMS@LIG composites behave as an orthogonally anisotropic conductive material, and we can conclude that $\rho_{33} = \rho_{11}$, $\rho_{13} = \rho_{31} = 0$, and $\rho_{23} = \rho_{32} = \rho_{12} = \rho_{21}$. Consequently, the resistivity tensor $\boldsymbol{\rho}$ ($\Omega\cdot\text{mm}$) for the anisotropic conductive PDMS@LIG composite can be expressed as:

$$\boldsymbol{\rho} = \begin{bmatrix} 12.367 & -1.138 & 0 \\ -1.138 & 17.751 & -1.138 \\ 0 & -1.138 & 12.367 \end{bmatrix} \tag{7}$$

Using $\boldsymbol{\sigma} = \boldsymbol{\rho}^{-1}$, the conductivity tensor $\boldsymbol{\sigma}$ ($\text{S}\cdot\text{mm}^{-1}$) can be expressed as:

$$\boldsymbol{\sigma} = \begin{bmatrix} 0.0813 & 0.00525 & 0.000483 \\ 0.00525 & 0.0570 & 0.00525 \\ 0.000483 & 0.00525 & 0.0813 \end{bmatrix} \tag{8}$$

To perform finite element simulations of the sensor's electromechanical behavior, it is essential to determine the relationship between resistance change and strain throughout the loading process. For piezoresistive sensors, resistance change under load is typically attributed to two factors: the geometric effect and the piezoresistive

effect:[39, 64-66]

$$\Delta R/R = (1+2\upsilon)\varepsilon + \Delta\rho/\rho \tag{9}$$

Where, $(1+2\upsilon)\varepsilon$ represents the influence of geometric effect, and $\Delta\rho/\rho$ represents the contribution from the piezoresistivity change of the materials. However, Eq. (9) is derived from the total differential equation,[67] which assumes small deformations and linear response. As a result, it is only applicable to piezoresistive sensors under small strains, such as metal foil strain gauges, which exhibit a linear relationship between resistance change and strain. For sensors subjected to large strains, where nonlinear behavior becomes significant, this equation can lead to considerable errors. Therefore, it is necessary to derive an equation for resistance change that is applicable to sensors operating under large strains.

Fig. 4(e) illustrates an elastic resistive block with a longitudinal resistivity $\rho_0$, subjected to tensile force and deformed under the influence of Poisson's ratio. Its initial length, width, and height are $l$, $w$, $h$, respectively. The longitudinal resistance is $R_0$, and the equation is as follows:

$$R_0 = \frac{\rho_0 l}{wh} \tag{10}$$

Assuming a strain of $\varepsilon$ is applied in the longitudinal direction, the length, width, and height become $l_\varepsilon$, $w_\varepsilon$ and $h_\varepsilon$, respectively, and the longitudinal resistivity is $\rho_\varepsilon$. The longitudinal resistance can then be expressed as:

$$R_\varepsilon = \frac{\rho_\varepsilon l_\varepsilon}{w_\varepsilon h_\varepsilon} \tag{11}$$

It can be easily derived that: $l_\varepsilon = l(1+\varepsilon)$, the elongation ratio is $\lambda = 1+\varepsilon$, and the true strain in the longitudinal direction is given by:

$$\varepsilon_{true}^{l} = \ln(1+\varepsilon) \tag{12}$$

According to the Poisson effect, the true strain in the transverse direction is given by:

$$\varepsilon_{true}^{t} = -\upsilon \cdot \ln(1+\varepsilon) \tag{13}$$

Where $\upsilon$ is the Poisson's ratio of the elastic resistive block. Therefore, the nominal strain in the transverse direction is given by:

$$\varepsilon^t = e^{\varepsilon_{true}^t} - 1 \tag{14}$$

By substituting Eq. (13) into Eq. (14) and simplifying:

$$\varepsilon^t = (1+\varepsilon)^{-\upsilon} - 1 \tag{15}$$

Thus, the width of the resistive block is $w_\varepsilon = w(1+\varepsilon^t)$, and the height is $h_\varepsilon = h(1+\varepsilon^t)$. Substituting these into Eq. (11) and simplifying yields:

$$R_\varepsilon = \frac{\rho_\varepsilon l}{wh(1+\varepsilon)^{-2\upsilon-1}} \tag{16}$$

By substituting Eq. (10) and Eq. (16) into the expression for normalized resistance response $\Delta R/R_0 = R/R_0 - 1$, and we can get:

$$\frac{\Delta R}{R_0} = \frac{\rho_\varepsilon}{\rho_0} \cdot \frac{1}{(1+\varepsilon)^{-2\upsilon-1}} - 1 \tag{17}$$

Conductivity $\sigma$ is the inverse of resistivity $\rho$, $\sigma = 1/\rho$. From Eq. (17), the change in conductivity is given by:

$$\frac{\sigma_\varepsilon}{\sigma_0} = \frac{1}{1+\frac{\Delta R}{R_0}} \cdot \frac{1}{(1+\varepsilon)^{-2\upsilon-1}} \tag{18}$$

Where $\sigma_0$ represents the initial conductivity in the longitudinal direction, and $\sigma_\varepsilon$ is the conductivity after strain $\varepsilon$ is applied. For PDMS, it is generally assumed that $\upsilon = 0.5$. The experimentally measured relationship between normalized resistance response and strain, shown in Fig. 4(c), is substituted into Eq. (18), yielding the curve of conductivity change with strain for the parallel specimen (the black curve in Fig. 4(f)). Eq. (S1) is a piecewise function employed to fit the curve. Evidently, this is the relationship between conductivity in the 1-direction $\sigma_{11}$ and strain. Following a similar derivation, the relationship between the conductivity change in the perpendicular specimen (i.e., the conductivity change in the 2-direction, $\sigma_{22}$) and the applied strain can be expressed as:

$$\frac{\sigma_\varepsilon}{\sigma_0} = \frac{1}{1+\frac{\Delta R}{R_0}} \cdot \frac{1}{(1+\varepsilon)} \tag{19}$$

The pink curve corresponding to specimen II in Fig. 4(e) illustrates the relationship between conductivity change in the 2-direction and strain, which is described by a cubic function of strain (Eq. S2).

As previously mentioned, the resistance response to strain in the thickness direction of PDMS@LIG cannot be directly measured due to its thinness. However, since the structural features in the thickness direction are analogous to those in the laser-scribed direction, it is assumed they exhibit similar electromechanical behavior. According to Eq. (15), the geometric deformation in the thickness direction (3-direction) under tensile strain is identical to that in the 2-direction. Consequently, by aligning the laser-scribing direction perpendicular to the tensile direction, a perpendicular specimen was prepared to obtain the relationship between conductivity in the thickness direction ($\sigma_{33}$) and tensile strain, as illustrated in Fig. S2. Regarding the relationship between the off-diagonal components and strain, experimental tests show that $\sigma_{12}$ remains essentially unchanged with increasing strain. For computational simplicity, the off-diagonal components are treated as constant.

**3.6 Finite element analysis (FEA) of PDMS@LIG sensors**

3D models were constructed using the steady-state thermomechanical module in ABAQUS software to simulate the mechanical properties and electrical response of the PDMS@LIG sensor. This method was chosen because the thermal and electrical behaviors are governed by the same equations,[50, 52, 68, 69] allowing for a unified simulation approach. The correspondence between the electrical property parameters and their thermal equivalents, including units, is shown in Table S2. Hybrid, linear, and three-dimensional brick elements (C3D8H) were used for the PDMS material, while the PDMS@LIG composites was meshed using eight-node hexahedral hybrid element (C3D8HT) from the coupled temperature-displacement element type. The analysis was carried out in two steps: first, an electric potential (with a voltage difference of 10V) was applied at the locations corresponding to the bonded wire electrodes; second, the tensile process was simulated, during which the change in total heat flux through the surface (SOH, representing current) at the electrodes was obtained. As analyzed in **Section 3.5**, the relationship between the change in conductivity of the PDMS@LIG composites and axial strain during sensor deformation has been formulated. Consequently, the conductivity vector $\mathbf{k}$ can be expressed as a function of axial strain ($\varepsilon_{11}$):

$$\mathbf{k}=\mathbf{k}_0 \times f(\varepsilon_{11}) \tag{20}$$

Where, $\mathbf{k}_0$ represents original conductivity vector. According to Fourier's law, the heat flux vector $\mathbf{f}$ can be written as:[69]

$$\mathbf{f} = -\mathbf{k} \cdot \mathbf{g} \tag{21}$$

with $\mathbf{g}$ as the spatial gradients of temperature, and the heat flux vector with respect to the spatial gradients of temperature can be written as

$$\frac{\partial \mathbf{f}}{\partial \mathbf{g}} = -\mathbf{k} \tag{22}$$

The electrical behavior of the PDMS@LIG composites is modeled using the user-defined subroutine UMATHT for material thermal behavior, while the user subroutine USDFLD is employed to obtain and transfer strain field variables to UMATHT at material points during the deformation of the sensor, thereby capturing the effect of strain on conductivity.

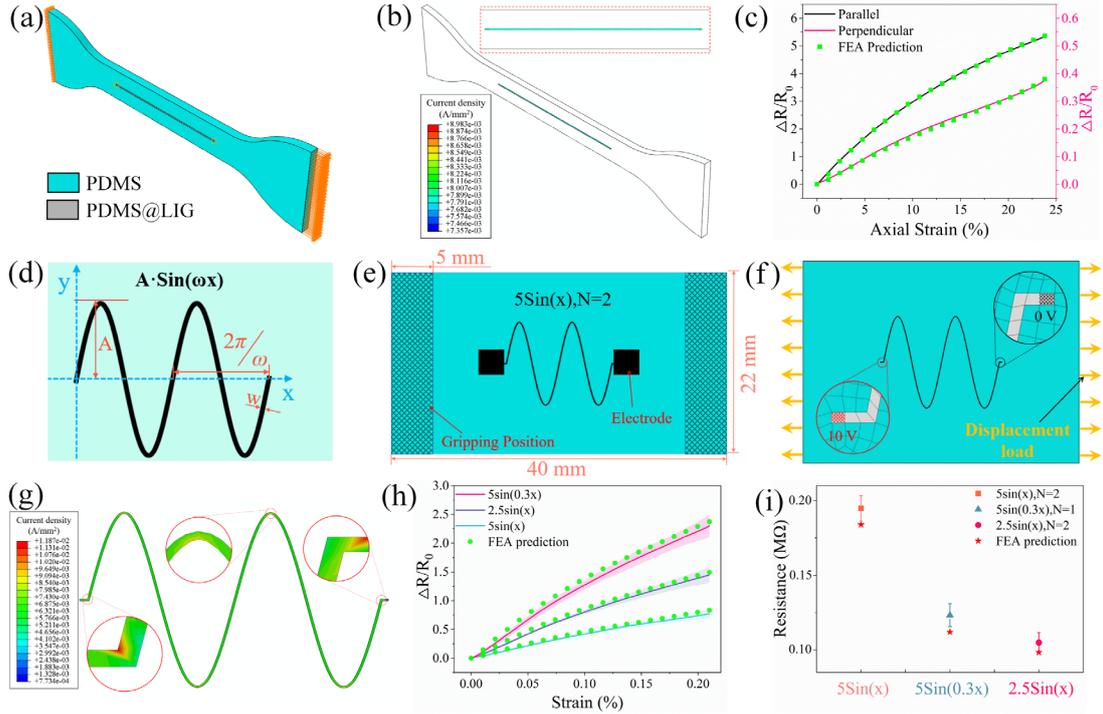

Fig. 5. (a) The parallel specimen model with applied loads. (b) Current density contour. (c) FEA predictions and experimental data of parallel and perpendicular specimens. (d) Schematic of the sinusoidal pattern design. (e) Planar schematic of the sinusoidal-patterned PDMS@LIG sensor. (f) Schematic of the boundary conditions applied to FEA model. (g) Current density contour of the PDMS@LIG composites under a 6 mm displacement load. (h) Comparison between the FEA predicted electromechanical response and experimental data for typical sinusoidal-patterned samples. (i) Comparison of FEA predicted resistance values with experimental measurements

Electromechanical response simulations were initially conducted for both parallel and perpendicular specimens. Fig. 5(a) presents the parallel specimen model with applied voltage and displacement loads, while Fig. S3 provides further comprehensive details of both the parallel and perpendicular specimen models. The strain distribution contour at the maximum displacement load of 10 mm is presented in Fig. S4, demonstrates a high degree of uniformity across the narrow section, with no evident strain concentrations. This uniformity is critical for accurately predicting the electromechanical relationship. Similarly, the electric field distribution depicted in Fig.

S5, reveals a consistent potential gradient along the axial direction, further contributing to the reliability of the simulation. Fig. 5(b) displays the current density contour, with an inset offering a magnified view of the PDMS@LIG section. The current density is highly uniform, a direct consequence of the homogeneous strain and electric field distributions. The comparison between the FEA predictions and experimental data is presented in Fig. 5(c), where the two data sets demonstrate a strong correlation ($R^2 >$ 0.99), confirming the accuracy of the derived conductivity-strain relationship and validating the effectiveness of the simulation method in modeling the sensor's electromechanical behavior.

Additionally, we predicted the behavior of the PDMS@LIG sensor with a sinusoidal patterned design, as illustrated in Fig. 5(d). The curve's shape is defined by key parameters such as amplitude (A), period ($\omega$), line width (w), and the number of cycles (N). Fig. 5(e) presents a planar schematic of a sensor with the sinusoidal patterned PDMS@LIG composites described by the function $5\sin(x)$ and two cycles (N=2). The sensor has a rectangular substrate with dimensions of $40 \times 22 \times 1$ mm, with 5 mm at each end reserved for clamping. Fig. S6 presents a photograph of the sensor, while Fig. S7 demonstrates its flexibility. The boundary conditions for the sinusoidal pattern sensor model are depicted in Fig. 5(f), where a voltage of 0V and 10V is applied at the electrode positions on either side of the PDMS@LIG composites, and displacement loads of 6 mm are applied to the ends of the substrate. The $5\sin(x)$ patterned PDMS@LIG sensor model is presented in Fig. S8, with the thickness of the PDMS@LIG composites section set to 0.03 mm, as determined from SEM. The potential distribution of the PDMS@LIG composites section is shown in Fig. S9, where it is observed to decrease uniformly along the sinusoidal curve. Fig. 5(g) shows the current density contour of the PDMS@LIG composites section under a 6 mm displacement load. The current density remains relatively uniform across most regions, except at the corners, where uneven distribution occurs due to the increased length—and consequently, higher resistance—of the outer edge of the bend. A magnified view of the bent region shows that the current on the inner side is approximately 1.5 times that on the outer side. Simulations and experimental tests were conducted to analyze the electromechanical behavior of PDMS@LIG sensors with three sinusoidal pattern parameters: 5sin(x), 2.5sin(x), and 5sin(0.3x). As shown in Fig. 5(h), the FEA predictions align closely with the experimental results. The accuracy of the finite element method was evaluated by calculating normalized chi-square values between the predicted and experimental data, yielding values of 1.22, 1.95, and 1.36, respectively. The initial resistance values of these sensors were also predicted, and a comparison with the experimental measurements is provided in Fig. 5(i). It is evident that the predicted values are generally lower than the measured ones. This discrepancy likely stems from the fact that the initial conductivity values used in the simulations were derived from tests on straight-shaped PDMS@LIG sensors, which experience less damage during the transfer process due to their simpler geometry. In contrast, the sinusoidal-patterned LIG may suffer more damage at the bends during transfer, resulting in higher measured resistance. Although the finite element method dose not perfectly predict the actual resistance values, particularly in capturing relative resistance differences among various specimens, which are accurately reflected by the model.

### 3.7 Parameter optimization

After validating the feasibility and accuracy of the finite element method, it can now be applied to rapidly optimize the sensor's pattern parameters. As the effect of line thickness is not considered in this study, the sinusoidal pattern design focuses on three key parameters: amplitude (A), period (ω), and the number of cycles (N). For practical implementation, the amplitude (A) is constrained between 2 and 8, the period (ω) between 0.1 and 10, and the number of cycles (N) between 1 and 5. Additionally, due to the limitations of the substrate's dimensions, these parameters must satisfy the following constraints:

$$8 \leq \frac{2N \cdot \pi}{\omega} \leq 24 \tag{23}$$

Where, setting $\frac{2N \cdot \pi}{\omega}$ to less than 24 rather than the total length 30 is to allocate space for the electrodes.

Within the specified parameter constraints, 50 sets of shape parameters were generated using the Latin hypercube sampling method. To reduce the significant workload associated with repeated modeling and computation, Python scripts were developed in the ABAQUS Scripting Interface for rapid modeling. A custom script was created for the sinusoidal-patterned sensor, with amplitude (A), period (ω), and number of cycles (N) as user-defined input parameters. During the simulation process, the script automatically retrieves each parameter sets from the pre-generated group, constructs the corresponding models, and submits them for computation.

Fig. 6(a) illustrates the normalized resistance response curves for these 50 parameter sets. The electromechanical behavior of these sensors was analyzed to determine the linearity and GF for each parameter set. Using these two primary performance metrics, Pareto front solutions were determined, as illustrated in Fig. 6(b). The analysis revealed that both the linearity and GF of the sensor exhibit a generally monotonic relationship with the product of amplitude (A) and period (ω)— $A \cdot \omega$. Therefore, $A \cdot \omega$ was used as the independent variable, and the two optimization objectives served as dependent variables to generate the plot shown in Fig. 6(c). After comprehensive evaluation, the optimal solution circled in green was selected, corresponding to A=7.12, ω=0.55, and N=2. Fig. 6(d) compares the electromechanical responses of the optimal patterned sample with the 5sin(0.3x) sample. Despite a slight reduction in sensitivity, the linearity $R^2$ increased from 0.982 to 0.990, leading to an overall optimization of the sensor's performance.

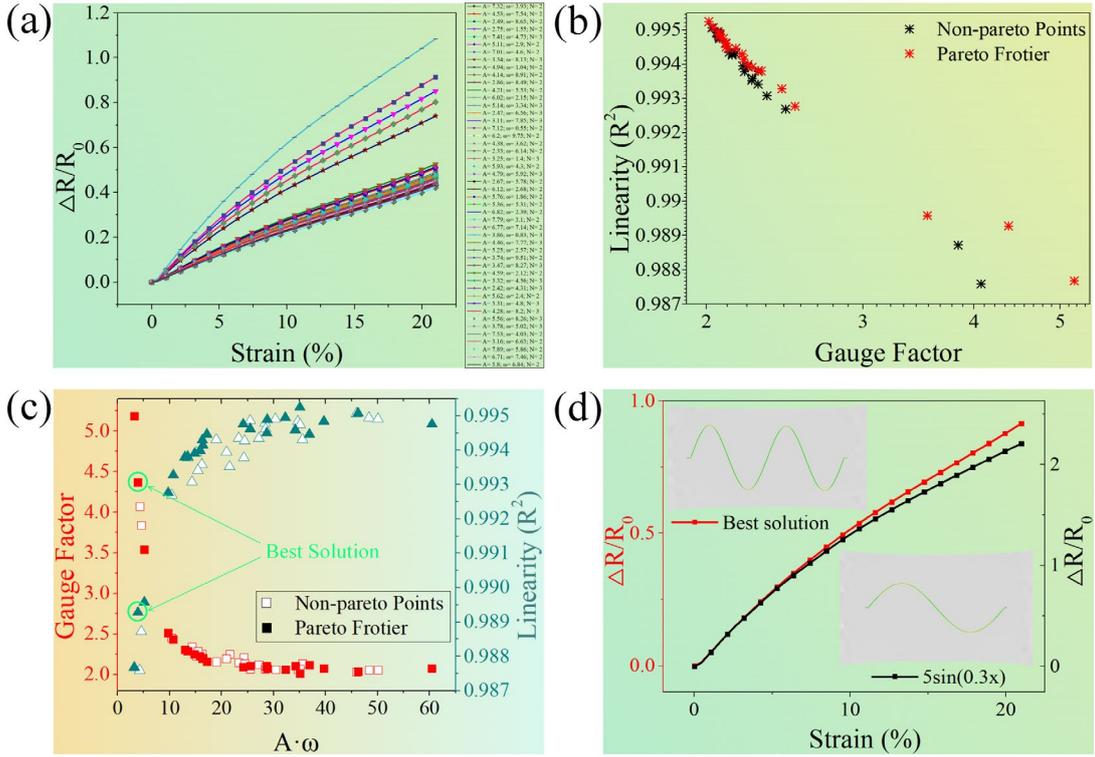

Fig. 6. (a) Normalized resistance response curves of 50 parameter sets. (b) Pareto front solutions. (c) Optimal Pareto front solutions. (d) Comparison of the electromechanical response between optimal and unoptimized samples.

## 4. Conclusion

In this study, we highlighted the significance of patterned design in flexible sensors, particularly focusing on the challenges of low efficiency and high costs in their development. Initially, PDMS@LIG sensors were developed using laser-induced graphene technology, and their piezoresistive properties with anisotropic conductivity were analyzed. We proposed a universal piezoresistive model for strain sensors, overcoming the limitations of earlier models that relied on total differentiation and were limited to small deformations and linear electromechanical responses. The electromechanical behavior of PDMS@LIG composites, derived from this piezoresistive model, was implemented in ABAQUS using the user-defined subroutine UMATHT for thermal behavior. The predicted sensing responses of both straight and sinusoidal LIG sensors showed good agreement with experimental data, validating the feasibility and accuracy of the piezoresistive model and finite element method. We then generated 50 sets of sinusoidal pattern parameters using the Latin hypercube sampling method and performed simulations of the corresponding sensors' electromechanical behavior. Using linearity and sensitivity as the optimization objectives, Pareto front solutions were obtained, and one was chosen as the optimal solution. Compared to the unoptimized sensor, the linearity $R^2$ improved by 0.008.

In conclusion, we presented a universal piezoresistive model and a comprehensive method for optimizing the shape parameters of patterned sensors using finite element analysis. This approach offers valuable guidance for the high-quality development of flexible sensors.

# References

[1] Guo X, Li Y, Zeng Z, Zhao Y, Lei X, Wang Y, et al. Ultra-sensitive flexible pressure sensor with hierarchical structural laser-induced carbon nanosheets/carbon nanotubes composite film. Compos Sci Technol. 2023;244:110290.

[2] Wang X, Liu X, Ge X, Schubert DW. Superior sensitive, high-tensile flexible fabric film strain sensor. Compos. Part A Appl. Sci. Manuf. 2023;172:107610.

[3] Zhu S, Lu Y, Wang S, Sun H, Yue Y, Xu X, et al. Interface design of stretchable and environment-tolerant strain sensors with hierarchical nanocellulose-supported graphene nanocomplexes. Compos. Part A Appl. Sci. Manuf. 2023;164:107313.

[4] Shang J-C, Yang H, Hong G-Q, Zhao W-H, Yang Y-F. Flexible pressure sensor enhanced by polydimethylsiloxane and microstructured conductive networks with positive resistance-pressure response and wide working range. Compos. B. Eng. 2023;264:110931.

[5] Hu T, Sheng B. A highly sensitive strain sensor with wide linear sensing range prepared on a hybrid-structured cnt/ecoflex film via local regulation of strain distribution. ACS Appl. Mater. Interfaces. 2024;16(16):21061–72.

[6] Gu Y, Zhang Z, Fan F, Wei L, Wu T, Wang D, et al. Designable high-performance TPU foam strain sensors towards human-machine interfaces. Compos. Part A Appl. Sci. Manuf. 2024;182:108169.

[7] Yang X, Chen W, Fan Q, Chen J, Chen Y, Lai F, et al. Electronic skin for health monitoring systems: properties, functions, and applications. Adv Mater. 2024;36(31):e2402542.

[8] Herbert R, Lim H-R, Yeo W-H. Printed, Soft, Nanostructured strain sensors for monitoring of structural health and human physiology. ACS Appl. Mater. Interfaces. 2020;12(22):25020-30.

[9] Zhao G, Sun J, Zhang M, Guo S, Wang X, Li J, et al. Highly strain-stable intrinsically stretchable olfactory sensors for imperceptible health monitoring. Adv Sci. 2023;10(29):2302974.

[10] Ma ZL, Wei AJ, Li YT, Shao L, Zhang HM, Xiang XL, et al. Lightweight, flexible and highly sensitive segregated microcellular nanocomposite piezoresistive sensors for human motion detection. Compos Sci Technol. 2021;203:108571.

[11] Ye Z, Zhao B, Wang Q, Chen K, Su M, Xia Z, et al. Crack-induced superelastic, strength-tunable carbon nanotube sponges. Adv. Funct. Mater. 2023;33(44):2303475.

[12] Sun Y, Liu K, Bu F, Meng R, Xie G, Guo K, et al. Low-cost, reliable and flexible piezoresistive pressure sensors coated with single layer graphene and silver nanowires on three-dimensional polyurethane sponge. Sens. Actuator A Phys. 2024;375:115524.

[13] Li Z, Wang S, Wang W, Wu J, Zhang Z, Li D, et al. Nacre-inspired MMT-MXene integrated shear-stiffening gel composites for personal safeguard and multi-functional electronics. Compos. B. Eng. 2024;280:111526.

[14] Liu M, Sheng Y, Huang C, Zhou Y, Jiang L, Tian M, et al. Highly stretchable and sensitive sbs/gr/cnts fibers with hierarchical structure for strain sensors. Compos.

Part A Appl. Sci. Manuf. 2023;164:107296.

[15] Ding Y, Dong H, Cao J, Zhang Z, Chen R, Wang Y, et al. A polyester/spandex blend fabrics-based e-textile for strain sensor, joule heater and energy storage applications. Compos. Part A Appl. Sci. Manuf. 2023;175: 107779.

[16] Lu H, Feng Y, Wang S, Liu J, Han Q, Meng Q. A high-performance, sensitive, low-cost lig/pdms strain sensor for impact damage monitoring and localization in composite structures. Nanotechnology. 2024;35(35):355702.

[17] He S, Wu J, Liu S, Wei Q, Liu S, Su B, et al. A fully integrated multifunctional flexible sensor based on nitrile rubber/carbon nanotubes/graphene composites for smart tire. Chem. Eng. J. 2024;486:150104.

[18] Ou C, Jiang H, Xiao L, Zhang D, Ma Y, Feng S, et al. Silicone/broadleaf wood fiber/mwcnts composite stretchable strain sensor for smart object identification. Sens. Actuator A Phys. 2023;364:114846.

[19] Ji J, Zhao W, Wang Y, Li Q, Wang G. Templated laser-induced-graphene-based tactile sensors enable wearable health monitoring and texture recognition via deep neural network. ACS Nano. 2023;17(20):20153-66.

[20] Davoodi E, Fayazfar H, Liravi F, Jabari E, Toyserkani E. Drop-on-demand high-speed 3D printing of flexible milled carbon fiber/silicone composite sensors for wearable biomonitoring devices. Addit. Manuf. 2020;32:101016.

[21] Zhu Y, Li Y, Xie D, Yan B, Wu Y, Zhang Y, et al. High-performance flexible tactile sensor enabled by multi-contact mechanism for normal and shear force measurement. Nano Energy. 2023;117:108862.

[22] Fan J, Kuo Y-C, Yin T, Guan P, Meng L, Chen F, et al. One-step synthesis of graphene-covered silver nanowires with enhanced stability for heating and strain sensing. ACS Appl. Mater. Interfaces. 2024;16(30):39600-12.

[23] Mao L, Pan T, Lin L, Ke Y, Su H, Li Y, et al. Simultaneously enhancing sensitivity and operation range of flexible pressure sensor by constructing a magnetic-guided microstructure in laser-induced graphene composite. Chem. Eng. J. 2024;481:148639.

[24] Hassan G, Bae J, Hassan A, Ali S, Lee CH, Choi Y. Ink-jet printed stretchable strain sensor based on graphene/zno composite on micro-random ridged pdms substrate. Compos. Part A Appl. Sci. Manuf. 2018;107:519-28.

[25] Shang J, Yang H, Yao X, Chen H. Structure driven piezoresistive performance design for rubbery composites-based sensors and application prospect: a review. Acta Mech Sin. 2024;40:423211

[26] Yang H, Shang J-C, Wang W-F, Yang Y-F, Yuan Y-N, Lei H-S, et al. Polyurethane sponges-based ultrasensitive pressure sensor via bioinspired microstructure generated by pre-strain strategy. Compos Sci Technol. 2022;221:109308.

[27] Wang Y, Chen Z, Mei D, Zhu L, Wang S, Fu X. Highly sensitive and flexible tactile sensor with truncated pyramid-shaped porous graphene/silicone rubber composites for human motion detection. Compos Sci Technol. 2022;217:109078.

[28] Lv Y, Zhang M, Zhao B, Qin Z, Chen K, Liu Y, et al. Flexible laser-reduced graphene with gradient-wrinkled microstructures for piezoresistive pressure sensors. ACS Appl. Nano Mater. 2024;7(16):18986-94.

[29] Chen J, Xia X, Yan X, Wang W, Yang X, Pang J, et al. Machine learning-enhanced biomass pressure sensor with embedded wrinkle structures created by surface buckling. ACS Appl. Mater. Interfaces. 2023;15(39):46440-8.

[30] Weng MC, Sun LQ, Qu SX, Chen LZ. Fingerprint-inspired graphene pressure sensor with wrinkled structure. Extreme Mech. Lett. 2020;37:100714.

[31] Liu Q, Chen J, Li Y, Shi G. High-performance strain sensors with fish-scale-like graphene-sensing layers for full-range detection of human motions. ACS Nano. 2016;10(8):7901-6.

[32] Wang W, Lu L, Li Z, Lin L, Liang Z, Lu X, et al. Fingerprint-inspired strain sensor with balanced sensitivity and strain range using laser-induced graphene. ACS Appl. Mater. Interfaces. 2021;14(1):1315-25.

[33] Yao HB, Ge J, Wang CF, Wang X, Hu W, Zheng ZJ, et al. A flexible and highly pressure-sensitive graphene-polyurethane sponge based on fractured microstructure design. Adv Mater. 2013;25(46):6692-8.

[34] Pang Y, Zhang K, Yang Z, Jiang S, Ju Z, Li Y, et al. Epidermis microstructure inspired graphene pressure sensor with random distributed spinosum for high sensitivity and large linearity. ACS Nano. 2018;12(3):2346-54.

[35] Huang B, Feng J, He J, Huang W, Huang J, Yang S, et al. High sensitivity and wide linear range flexible piezoresistive pressure sensor with microspheres as spacers for pronunciation recognition. ACS Appl. Mater. Interfaces. 2024;16(15):19298-308.

[36] Xu J, Zhang L, Lai X, Zeng X, Li H. Wearable rgo/mxene piezoresistive pressure sensors with hierarchical microspines for detecting human motion. ACS Appl. Mater. Interfaces. 2022;14:27262−73.

[37] Li G, Chen D, Li C, Liu W, Liu H. Engineered microstructure derived hierarchical deformation of flexible pressure sensor induces a supersensitive piezoresistive property in broad pressure range. Adv Sci. 2020;7(18):2000154.

[38] Mousavi S, Howard D, Zhang F, Leng J, Wang CH. Direct 3d printing of highly anisotropic, flexible, constriction-resistive sensors for multidirectional proprioception in soft robots. ACS Appl. Mater. Interfaces. 2020;12(13):15631-43.

[39] Pimentel E, Costa P, Tubio CR, Vilaça JL, Costa CM, Lanceros-Méndez S, et al. Printable piezoresistive polymer composites for self-sensing medical catheter device applications. Compos Sci Technol. 2023;239:110071.

[40] Chen X, Wang F, Shu L, Tao X, Wei L, Xu X, et al. A Single-material-printed, Low-cost design for a Carbon-based fabric strain sensor. Mater. Des. 2022;221:110926.

[41] Li L, Gao M, Guo Y, Sun J, Li Y, Li F, et al. Transparent ag@au–graphene patterns with conductive stability via inkjet printing. J. Mater. Chem. C. 2017;5(11):2800-6.

[42] Shahariar H, Kim I, Soewardiman H, Jur JS. Inkjet printing of reactive silver ink on textiles. ACS Appl. Mater. Interfaces. 2019;11(6):6208-16.

[43] Chen X, Yang X, Xia X, Zhao J. A highly responsive and sensitive flexible strain sensor with conductive cross-linked network by laser direct writing. Sens. Actuator A Phys. 2024;378:115809.

[44] Tang L, Zhou J, Zhang D, Sheng B. Laser-induced graphene electrodes on poly(ether–ether–ketone)/pdms composite films for flexible strain and humidity sensors. ACS Appl. Nano Mater. 2023;6(19):17802-13.

[45] Zhu J, Xiao Y, Zhang X, Tong Y, Li J, Meng K, et al. Direct laser processing and functionalizing pi/pdms composites for an on-demand, programmable, recyclable device platform. Adv Mater. 2024;36(35):e2400236.

[46] Xu K, Lu Y, Honda S, Arie T, Akita S, Takei K. Highly stable kirigami-structured stretchable strain sensors for perdurable wearable electronics. J. Mater. Chem. C. 2019;7(31):9609-17.

[47] Yang Y, Wang H, Hou Y, Nan S, Di Y, Dai Y, et al. Mwcnts/pdms composite enabled printed flexible omnidirectional strain sensors for wearable electronics. Compos Sci Technol. 2022;226:109518.

[48] Peng S, Wu S, Yu Y, Sha Z, Li G, Hoang TT, et al. Carbon nanofiber-reinforced strain sensors with high breathability and anisotropic sensitivity. J. Mater. Chem. A. 2021;9(47):26788-99.

[49] Arana G, Gamboa F, Avilés F. Piezoresistive and thermoresistive responses of carbon nanotube-based strain gauges with different grid geometric parameters. Sens. Actuator A Phys. 2023;359:114477.

[50] Yang H, Yuan L, Yao X, Fang D. Piezoresistive response of graphene rubber composites considering the tunneling effect. J. Mech. Phys. Solids. 2020;139:103943.

[51] Önder MN, Gülgün MA, Papila M. Piezoresistivity analyses of gnp-filled composite piezoresistor under cycling loading and correlation with the monte carlo percolation model. Compos Sci Technol. 2024;254:110641.

[52] Lu X, Yvonnet J, Detrez F, Bai J. Multiscale modeling of nonlinear electric conductivity in graphene-reinforced nanocomposites taking into account tunnelling effect. J. Comput. Phys. 2017;337:116-31.

[53] Matos MAS, Tagarielli VL, Baiz-Villafranca PM, Pinho ST. Predictions of the electro-mechanical response of conductive cnt-polymer composites. J. Mech. Phys. Solids. 2018;114:84-96.

[54] Arana G, Mora A, Pérez I, Avilés F. Design and analysis of a carbon nanotube-based strain gauge via multiscale modeling. Meccanica. 2023;58(8):1717-32.

[55] Yang H, Yuan L, Yao X, Zheng Z, Fang D. Monotonic strain sensing behavior of self-assembled carbon nanotubes/graphene silicone rubber composites under cyclic loading. Compos Sci Technol. 2020;200:108474.

[56] Lin G, Zhou T, Zhou Z, Sun W. Laser induced graphene for emi shielding and ballistic impact damage detection in basalt fiber reinforced composites. Compos Sci Technol. 2023;242:110182.

[57] Luo S, Hoang PT, Liu T. Direct laser writing for creating porous graphitic structures and their use for flexible and highly sensitive sensor and sensor arrays. Carbon. 2016;96:522-31.

[58] Liu W, Huang Y, Peng Y, Walczak M, Wang D, Chen Q, et al. Stable wearable strain sensors on textiles by direct laser writing of graphene. ACS Appl. Nano Mater. 2020;3(1):283-93.

[59] Zhang B, Song J, Yang G, Han B. Large-scale production of high-quality graphene using glucose and ferric chloride. Chem Sci. 2014;5(12):4656-60.

[60] Huang L, Liu Y, Ji L-C, Xie Y-Q, Wang T, Shi W-Z. Pulsed laser assisted reduction of graphene oxide. Carbon. 2011;49(7):2431-6.

[61] Chen R, Liu S, Zhang C, Jiang C, Zhou W, Chen P, et al. Laser fabrication of humidity sensors on ethanol-soaked polyimide for fully contactless respiratory monitoring. ACS Appl. Mater. Interfaces. 2024;16(34):45252-64.

[62] Deng B, Wang Z, Liu W, Hu B. Multifunctional motion sensing enabled by laser-induced graphene. Materials. 2023;16(19):6363.

[63] Duy LX, Peng Z, Li Y, Zhang J, Ji Y, Tour JM. Laser-induced graphene fibers. Carbon. 2018;126:472-9.

[64] Guo Z, Xu J, Chen Y, Guo Z, Yu P, Liu Y, et al. High-sensitive and stretchable resistive strain gauges: parametric design and diw fabrication. Compos. Struct. 2019;223:110955.

[65] Li Q, Hong C, Xie H, Lai H, Shen Y, Xu S, et al. Piezoresistive behavior in cement-based sensors: Nonlinear modeling and preliminary application. Compos. Part A Appl. Sci. Manuf. 2023;175:107786.

[66] Lu Y, Liu Z, Yan H, Peng Q, Wang R, Barkey ME, et al. Ultrastretchable conductive polymer complex as a strain sensor with a repeatable autonomous self-healing ability. ACS Appl. Mater. Interfaces. 2019;11(22):20453-64.

[67] Beeby S. MEMS mechanical sensors: Artech House; 2004.

[68] Zhang J, Wang Z, Shang C, Qian Z, Wu Z, Yu X, et al. Piezoresistive relaxation and creep model of porous polymer nanocomposite supported by experimental data. Sens. Actuator A Phys. 2024;366:115002.

[69] Simulia DS. Abaqus 2024 Documentation. 2024.

# Supplementary Material for

# Geometric Optimization of Patterned Conductive Polymer Composite-based Strain Sensors Toward Enhanced Sensing Performance

**S1. Supplementary Figures.**

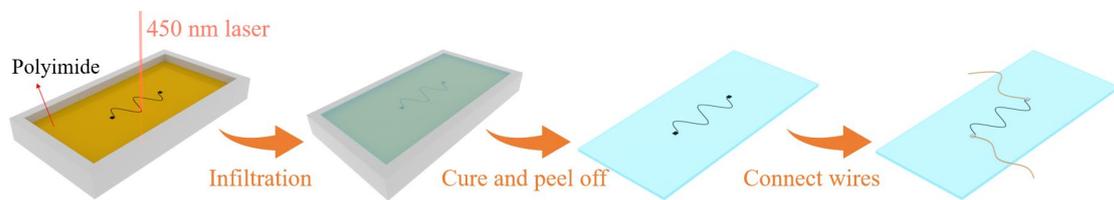

Fig S1. Fabrication diagrams of the patterned PDMS@LIG specimen

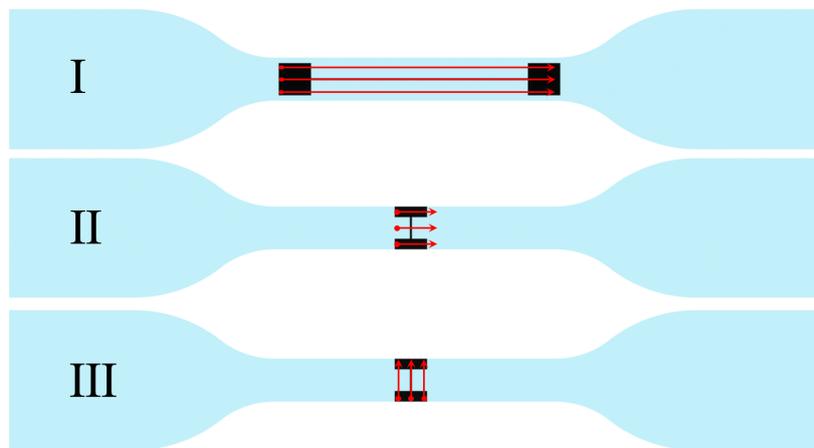

Fig S2. Relationship between the laser scribing direction and the specimens shown in Fig 4(c).

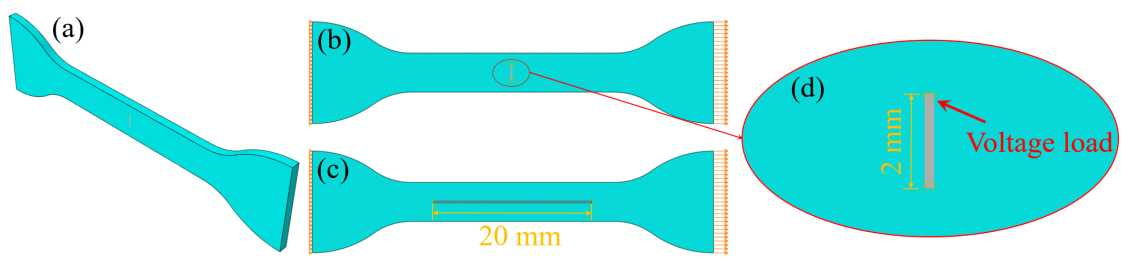

Fig S3. The parallel and perpendicular specimen models.

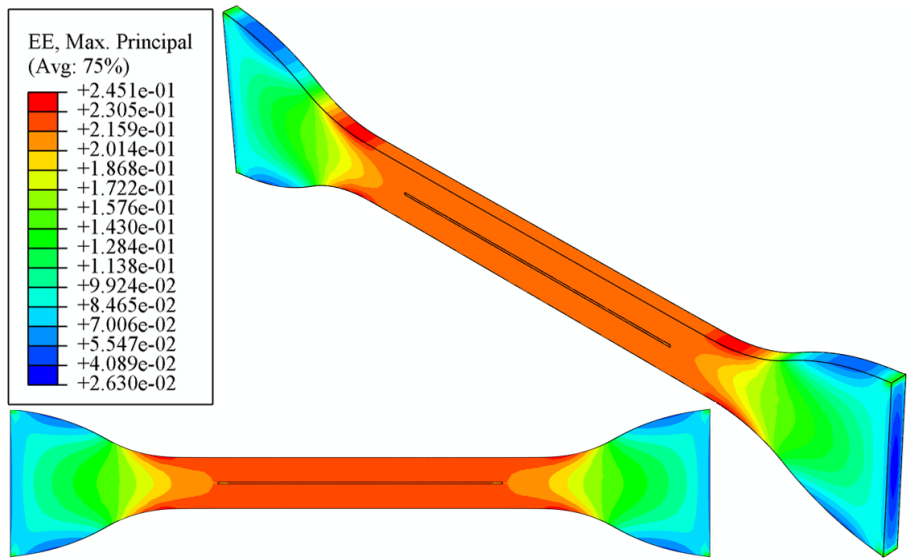

Fig S4. The strain distribution contour of the parallel sample.

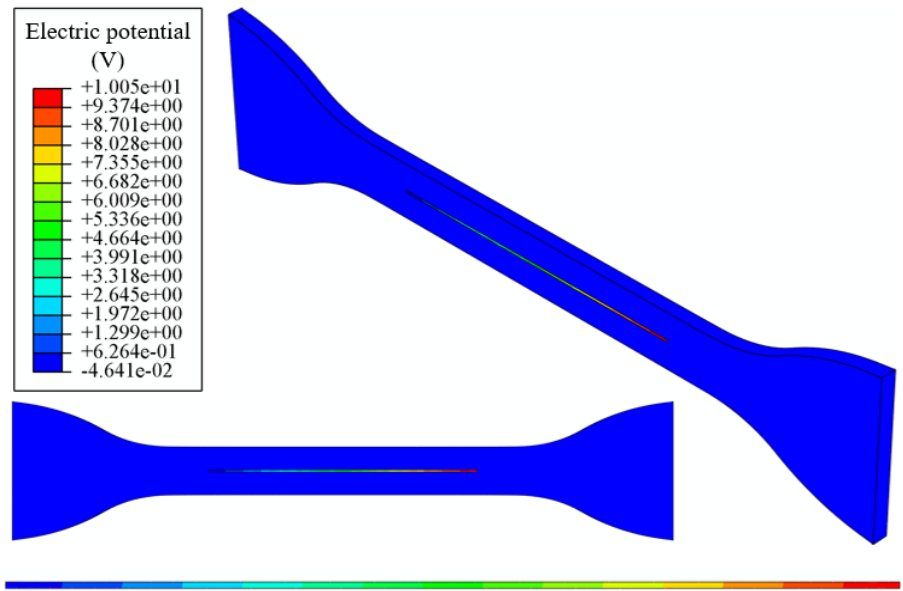

Fig S5. The potential contour of the parallel sample.

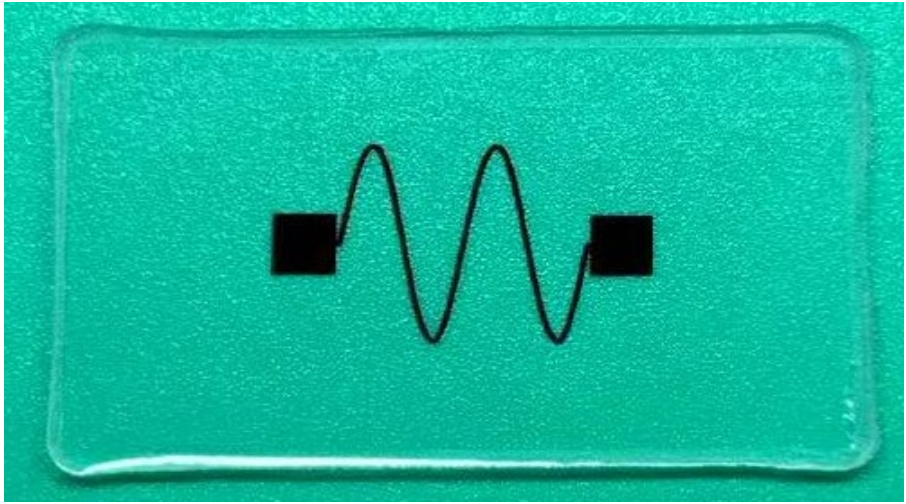

Fig S6. Photograph of the sinusoidal patterned sample.

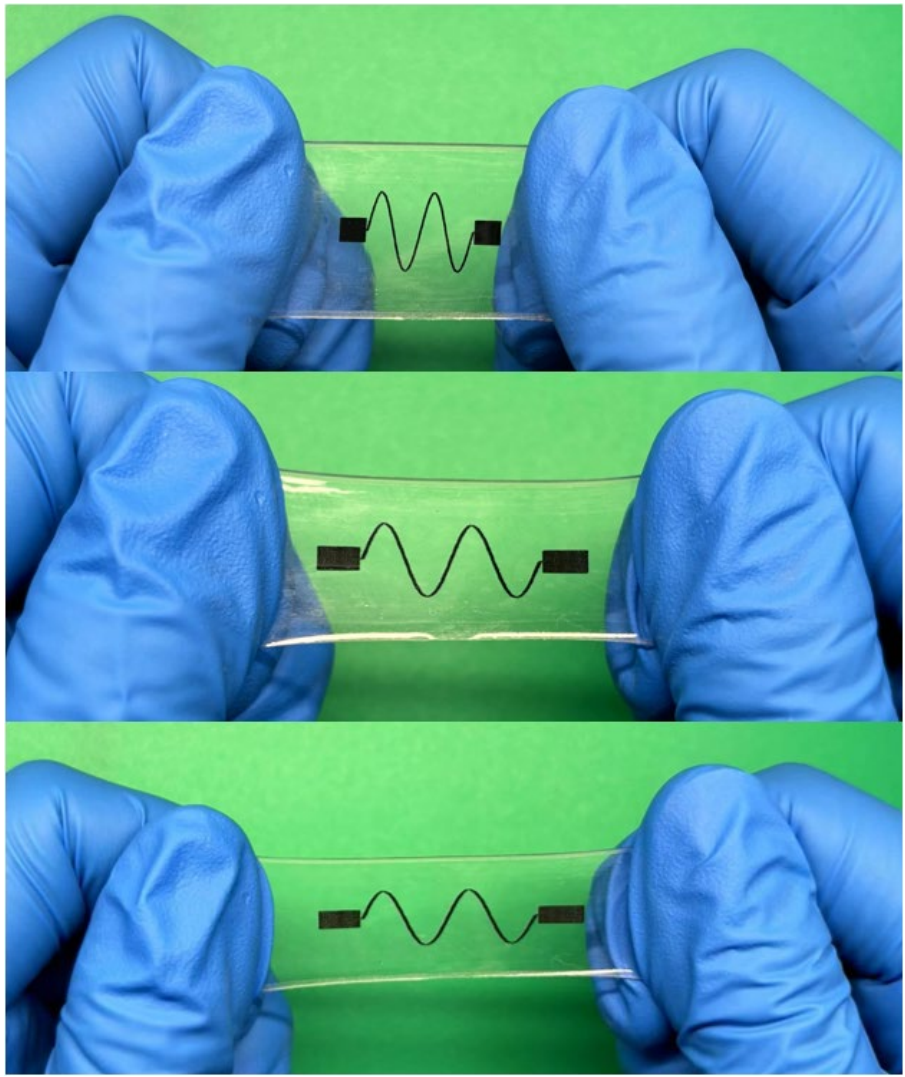

Fig S7. Demonstration of the sinusoidal patterned sensor's flexibility.

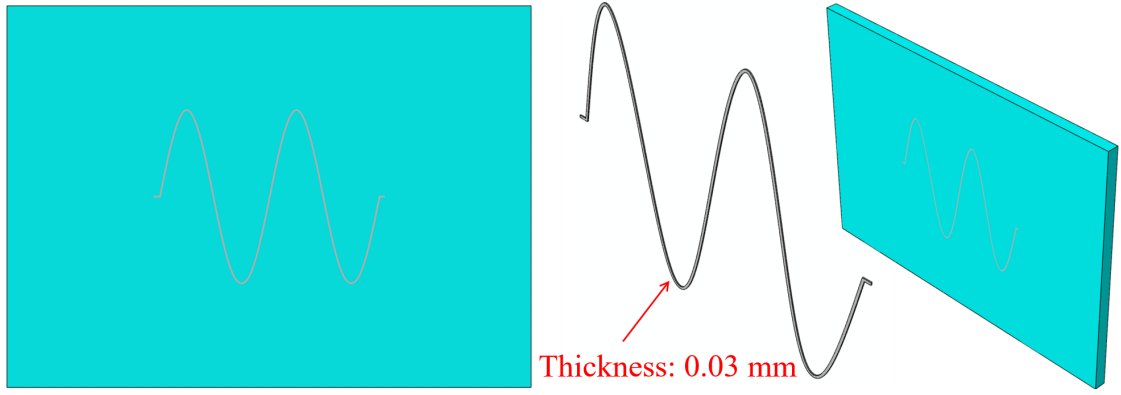

Fig S8. The sinusoidal patterned specimen models (5sin(x), N=2).

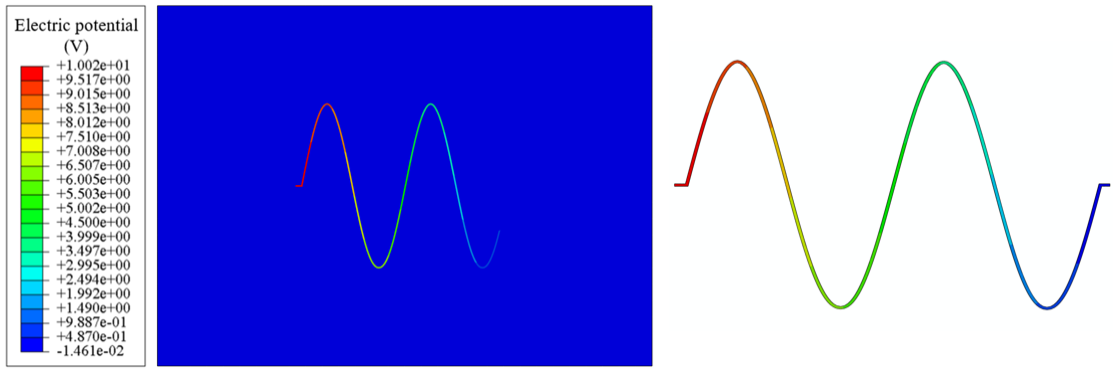

Fig S9. The potential contour of sinusoidal patterned specimen models (5sin(x), N=2).

## S2. The relationship between resistivity changes in various directions and principal strain.

Conductivity change with strain in the 1-direction:

$$\frac{\sigma_\varepsilon}{\sigma_0} = 0.7287 \times e^{-33.9668 \times \varepsilon_{11}} + 0.2713 \tag{S1}$$

Conductivity change with strain in the 2-direction:

$$\frac{\sigma_\varepsilon}{\sigma_0} = -18.5462 \times \varepsilon_{11}^3 + 10.4109 \times \varepsilon_{11}^2 - 3.2555 \times \varepsilon_{11} + 1.0 \tag{S2}$$

Conductivity change with strain in the 3-direction:

$$\frac{\sigma_\varepsilon}{\sigma_0} = 1.3898 \times \varepsilon_{11}^2 - 1.4626 \times \varepsilon_{11} + 1.0 \tag{S3}$$

**S3. Supplementary tables.**

Table S1. Laser Fluence and Corresponding Laser Parameters

| Laser fluence (J/cm$^2$) | Laser power (%) | Scanning speed (mm) |
|---|---|---|
| 5.7 | 15 | 262.79 |
| 7.66 | 15 | 195.81 |
| 10.2 | 20 | 195.81 |
| 12.62 | 20 | 158.51 |
| 15.85 | 25 | 158.51 |
| 17.0 | 27 | 158.51 |

Table S2. Corresponding relationship between electrical and thermal parameters

| Electrical physical quantities | Thermophysical quantities |
|---|---|
| Current density (A/mm$^2$) | Heat flux density (W/mm$^2$) |
| Potential (V) | Temperature (K) |
| Electrical conductivity (S/mm) | Heat conductivity (W·m$^{-1}$·K$^{-1}$) |